\documentclass[5p,longtitle]{elsarticle}
\usepackage{amsmath}
\usepackage{amssymb}

\usepackage[english]{babel}

\usepackage{amsfonts}
\usepackage{latexsym}
\usepackage[mathcal]{eucal}
\usepackage{subfigure}
\usepackage{graphicx}
\usepackage{bm}

\usepackage{hhline}

\usepackage{multirow}

\newcommand{\BBless}{\ensuremath{0\nu\beta\beta}~}

\newcommand{\elem}[2]{\ensuremath{{}^{#1}}#2}

\newcommand{\braket}[1]{\left(#1\right)}

\newcommand{\Tz}{\ensuremath{T^{0\nu}_{1/2}}~}

\newcommand{\TO}{TeO$_2$~}

\newcommand{\tect}{\elem{130}{Te}~}

\newcommand {\nmm}{\ensuremath{m_{\beta\beta}}}

\DeclareMathOperator{\erf}{erf}

\usepackage[usenames,dvipsnames]{color}
\definecolor{gelb}{rgb}{1,1,.4}
\definecolor{darkcyan}{rgb}{0,0.5,0.5}
\definecolor{black}{rgb}{0,0,0}
\definecolor{red}{rgb}{1,0,0}
\definecolor{blue}{rgb}{0,0,1}
\definecolor{ylw}{rgb}{0.8,0.8,0}
\definecolor{green}{rgb}{0,1,0}

\usepackage{marginnote}

\journal{Astroparticle Physics}

\begin{document}
\title{Sensitivity and Discovery Potential of CUORE to Neutrinoless Double-Beta Decay}

\author[INFNMilano]{F.~Alessandria}
\author[MilanoPoli]{R.~Ardito}
\author[USC,LNGS]{D.~R.~Artusa}
\author[USC]{F.~T.~Avignone~III}
\author[INFNLegnaro]{O.~Azzolini}
\author[LNGS]{M.~Balata}
\author[BerkeleyPhys,LBNLNucSci,LNGS]{T.~I.~Banks}
\author[INFNBologna]{G.~Bari}
\author[LBNLMatSci]{J.~Beeman}
\author[Roma,INFNRoma]{F.~Bellini}
% ...

\address{(CUORE Collaboration)}

\address[INFNMilano]{INFN - Sezione di Milano, Milano I-20133 - Italy}
\address[MilanoPoli]{Dipartimento di Ingegneria Strutturale, Politecnico di Milano, Milano I-20133 - Italy}
\address[USC]{Department of Physics and Astronomy, University of South Carolina, Columbia, SC 29208 - USA}
\address[LNGS]{INFN - Laboratori Nazionali del Gran Sasso, Assergi (L'Aquila) I-67010 - Italy}
\address[INFNLegnaro]{INFN - Laboratori Nazionali di Legnaro, Legnaro (Padova) I-35020 - Italy}
\address[BerkeleyPhys]{Department of Physics, University of California, Berkeley, CA 94720 - USA}
\address[LBNLNucSci]{Nuclear Science Division, Lawrence Berkeley National Laboratory, Berkeley, CA 94720 - USA}
\address[INFNBologna]{INFN - Sezione di Bologna, Bologna I-40127 - Italy}
\address[LBNLMatSci]{Materials Science Division, Lawrence Berkeley National Laboratory, Berkeley, CA 94720 - USA}
\address[Roma]{Dipartimento di Fisica, Sapienza Universit\`a di Roma, Roma I-00185 - Italy }
\address[INFNRoma]{INFN - Sezione di Roma, Roma I-00185 - Italy }
% ...

\fntext[fn1]{Deceased}
\fntext[fn2]{Presently at: Los Alamos National Laboratory, Los Alamos, NM 87545 - USA}
\fntext[fn3]{Presently at: Sandia National Laboratories, Livermore, CA 94551 - USA}


%

\begin{abstract}
  We present a study of the sensitivity and discovery potential of CUORE, a bolometric double-beta decay
  experiment under construction at the Laboratori Nazionali del Gran Sasso in Italy.
  Two approaches to the computation of experimental sensitivity for various background scenarios are presented, and an extension of the sensitivity formulation to the discovery potential case is also discussed. Assuming a background rate of
  $10^{-2}$~cts/(keV\,kg\,y), we find that, after 5 years of live time, CUORE will have a 1$\sigma$
  sensitivity to the neutrinoless double-beta decay half-life of 
  $\widehat{\Tz\!\!}(1\sigma) = 1.6 \times 10^{26}$~y and thus a potential to probe the effective
  Majorana neutrino mass down to
  40--100~meV; the sensitivity at 1.64$\sigma$, which corresponds to 90\%~C.L., will be $\widehat{\Tz\!\!}(1.64\sigma) = 9.5 \times 10^{25}$~y.  This range is compared with the claim of observation of
  neutrinoless double-beta decay in \elem{76}{Ge}
  and the preferred range in the neutrino mass parameter space from
  oscillation results.
\end{abstract}


\maketitle

%


\section{Introduction}


Neutrinoless double-beta decay (\BBless\!\!) (see Refs.~\cite{Elliott:2002xe,RevModPhys.80.481,Barabash:2008dj} for recent reviews) is a rare nuclear process hypothesized to occur if neutrinos are Majorana particles. In fact, the search for \BBless is currently the only experimentally feasible method to establish the Majorana nature of the neutrino. The observation of \BBless may also
probe the absolute mass of the neutrino and the neutrino mass hierarchy. Many experiments, focusing on several different candidate decay nuclides and utilizing various detector techniques, have sought evidence of this decay~\cite{Andreotti:2010vj,KlapdorKleingrothaus:2000sn,PhysRevD.65.092007,Bernabei200223,PhysRevLett.95.182302}; next-generation detectors are currently under development and construction and will begin data taking over the next few years.
Evidence of \BBless in $\rm ^{76}Ge$
has been reported~\cite{KlapdorKleingrothaus:2001ke,KlapdorKleingrothaus:2004ge,KlapdorKleingrothaus:2006ff} but has yet to be confirmed~\cite{Aalseth:2002dt,Harney:2001wb,Feruglio:2002af,Bakalyarov:2003jk}, and recent experimental results are in conflict with the claim~\cite{Bergstrom:2012nx}.

The Cryogenic Underground Observatory for Rare Events (CUORE)~\cite{Arnaboldi:2002du,Ardito:2005ar} is designed to search
for \BBless in \elem{130}{Te}.
Crystals made of natural $\rm TeO_{2}$, with an isotopic abundance of
$34.167\%$ of $\rm ^{130}Te$~\cite{Fehr:200483}, will be operated as bolometers, serving as source and detector at the
same time.  Such detectors combine excellent energy resolution with low intrinsic
background, and they have been operated in stable conditions
underground for several years~\cite{Alessandrello1992176,Alessandrello1995363,Alessandrello1998156}.  Individual detectors can be produced
with masses up to $\sim1$~kg, allowing for the construction of
close-packed large-mass arrays. Bolometric detectors enable precision measurement of the energy spectrum of events inside the crystals, allowing the search for an excess of events above background in a narrow window around the transition energy of the isotope of interest.  Such a peak constitutes the signature of \BBless\!\!, and if it is observed, the \BBless half-life can be determined from the number of observed events.

The current best limit on \BBless in $\rm ^{130}Te$ comes from the Cuoricino
experiment~\cite{Andreotti:2010vj,Arnaboldi:2005cg,Arnaboldi:2008ds}, which operated 58
crystals of natural \TO and 4 enriched \TO crystals (containing approximately $11$~kg of $^{130}$Te in total) in the Laboratori Nazionali del Gran Sasso, Italy, from 2003--2008.
With a total exposure of 19.75~kg\,y, Cuoricino set a limit of $\Tz > 2.8 \times 10^{24}$~y (90$\%$ C.L.)~\cite{Andreotti:2010vj} on the \BBless half-life of $^{130}$Te.

CUORE, the follow-up experiment to
Cuoricino, is currently under construction and will exploit the
experience and results gained from its predecessor. With its 988 detectors
and a mass of $\sim206$~kg of \elem{130}{Te}, CUORE will be larger by more than an order of magnitude. Background rates are also expected to be reduced
by approximately an order of magnitude with respect to Cuoricino.

In this study, we discuss the sensitivities of CUORE and of CUORE-0,
  a CUORE-like tower operated in the Cuoricino cryostat. We start by providing the detailed
assumptions and formulas for the sensitivity estimations. We then
review the experimental setup and parameters from which the
sensitivity values are calculated. Finally, we compare the sensitivities with the claim of observation of \BBless in \elem{76}{Ge} and the preferred range of neutrino
masses from oscillation results.

\section{Physics Reach of Double-Beta Decay Experiments}
\label{sec:sens}

After introducing some basic \BBless formulas in Sec.~\ref{sec:dbdphys}, we discuss several possible approaches to expressing the capabilities of a \BBless experiment in terms of the physics quantities it aims to explore.

\subsection{Observables of Double-Beta Decay}
\label{sec:dbdphys}

Double-beta decay is a second-order weak process, so half-lives are typically long: two-neutrino double-beta decay half-lives are at least of order $10^{18}$~years, while current limits on \BBless half-lives are
on the order of $10^{24}$~years or greater. With such long half-lives, the
radioactive decay law can be approximated as

\begin{equation}
N(t) \simeq N_{0} \braket{1- \ln(2)\cdot\frac{t}{T_{1/2}}},
\end{equation}
where $T_{1/2}$ is the half-life, $N_{0}$ is the initial number of atoms and $N(t)$ is the number of atoms left after time $t$ has passed.

Assuming that the exchange of a light Majorana neutrino is the dominant \BBless mechanism, the effective
Majorana mass of the electron neutrino can be inferred from the
\BBless half-life as follows~\cite{RevModPhys.80.481}:

\begin{equation}
\nmm = \frac{m_{e}}{\sqrt{F_{N}\cdot\Tz\!\!}},
\label{eq:mnu-NME}
\end{equation}
where $m_{e}$ is the electron mass, $F_N$ is a nuclear structure factor of merit that includes the nuclear
matrix elements (NME) and the phase space of the \BBless transition, and \Tz is the \BBless half-life.

The calculation of NMEs is difficult, and depending on the details of the underlying theoretical models, a range of values can be obtained, though there are indications of an underlying mechanism correlating NMEs and phase space factors that may allow a reduction of this theoretical uncertainty in future~\cite{Robertson:2013cy}. For the purpose of this work, we will
consider recent calculations from five different methods:
Quasiparticle Random Phase Approximation (QRPA) (carried out by two different
groups: the calculations of the T\"{u}bingen group are henceforth denoted by QRPA-T, and the calculations of the Jyv\"{a}skyl\"{a} group are henceforth denoted by QRPA-J), the Interacting
Shell Model (ISM), the Interacting Boson Model (IBM), the projected-Hartree-Fock-Bogoliubov model (PHFB), and the Generating Coordinate Method (GCM). $F_{N}$ values
and references are shown in Tab.~\ref{tab:nmev}. These values are
calculated using the NMEs reported by each group and the recent
phase space calculations of~\cite{Kotila:2012zza}, taking care to match the values of the axial vector coupling constant $g_{A}$ reported in each reference.  For most calculations, several values of the NMEs are reported
depending on the choice of the input parameters and assumptions in the model.
Therefore, we quote ranges of possible nuclear factors of merit,
taking the maximum and minimum values of the NME ranges reported in each case, with the exception of QRPA-T and PHFB, for which we include only the ranges arising from the use of the coupled-cluster method short-range correlations, following the indicated preferences of the authors. No statistical meaning
is implied in the use of these ranges.

\begin{table*}[tbp]
  \centering
  \caption{\BBless nuclear factors of merit $F_{N}$, as defined in Eq.~\eqref{eq:mnu-NME}, for the candidate \BBless nuclides discussed in this paper, according to different evaluation methods and authors. QRPA: Quasiparticle Random Phase Approximation; ISM: Interacting Shell Model; IBM: Interacting Boson Model; PHFB: projected-Hartree-Fock-Bogoliubov model; GCM: Generating Coordinate Method. See Sec.~\ref{sec:dbdphys} for details. The phase space values used in calculating $F_{N}$ values are taken from~\cite{Kotila:2012zza}.}
  \small
  \begin{tabular}{c@{\hspace{10mm}}c@{\hspace{5mm}}c@{\hspace{5mm}}c@{\hspace{5mm}}c@{\hspace{5mm}}c@{\hspace{5mm}}c}
    \noalign{\smallskip}
    \hline
    \noalign{\smallskip}
    \hline
    \noalign{\smallskip}
     & \multicolumn{6}{c}{\BBless nuclear factor of merit $F_{N}$} \\
     & \multicolumn{6}{c}{\scriptsize{($10^{-13}$ y$^{-1}$)}} \\
    Isotope & \footnotesize{QRPA-T~\cite{0954-3899-39-12-124006}} & \footnotesize{QRPA-J~\cite{0954-3899-39-12-124005}} & \footnotesize{ISM~\cite{Menendez:2008jp}} & \footnotesize{IBM-2~\cite{PhysRevC.87.014315}} & \footnotesize{PHFB~\cite{PhysRevC.82.064310}} & \footnotesize{GCM~\cite{PhysRevLett.105.252503}}\\ 
    \noalign{\smallskip}
    \hline
    \noalign{\smallskip}
    \tect & 3.56 -- 10.6 & 3.34 -- 9.28 & 1.56 -- 2.44 & 5.99 -- 7.84 & 3.10 -- 9.11 & 9.14 \\ 
    \elem{76}{Ge} & 1.15 -- 3.06 & 0.628 -- 1.89 & 0.305 -- 0.456 & 1.80 -- 2.33 & --- & 1.22 \\
    \noalign{\smallskip}
    \hline
    \noalign{\smallskip}
    \hline
  \end{tabular}
  \label{tab:nmev}
\end{table*}

\subsection{Sensitivity with Respect to Background Fluctuation}
\label{sec:back-fluct}

The mean value $S_{0}$ of the \BBless signal, i.e., the expected number of \BBless decays interacting in the detector during the live time $t$, is

\begin{equation}
S_{0} = \frac{M\cdot N_A\cdot a \cdot \eta}{W} \cdot \ln(2) \cdot \frac{t}{\Tz\!\!} \cdot \varepsilon,
\label{eq:nbb-full}
\end{equation}
where $M$ is the total active mass, $\eta$ is the stoichiometric coefficient of the \BBless candidate (i.e., the number of nuclei of the candidate \BBless element per molecule of the active mass), $W$ is the molecular weight of the active mass, $N_{A}$ is the Avogadro constant, $a$ is the isotopic abundance of the candidate \BBless nuclide and $\varepsilon$ is the physical detector efficiency.

In Eq.~\eqref{eq:nbb-full}, \Tz refers to the (unknown) \emph{true value} of the \BBless half-life, and $S_{0}$ is therefore also unknown. The background-fluctuation sensitivity formulates the sensitivity in reference to the magnitude of the observed-count fluctuations due to background expected in an experiment. In our derivation, we first determine sensitivity in terms of a number of counts (analogous to $S_{0}$) and then use Eq.~\eqref{eq:nbb-full} to convert to a half-life sensitivity (analogous to \Tz\!\!). In order to prevent confusion between sensitivities and true values, hatted quantities (e.g., $\widehat{\Tz\!\!}$, $\widehat{S_{0}}$) will be used to represent the sensitivities corresponding to the unhatted true values.

An experiment can expect to see a background contribution to the counts acquired in the energy window of interest for the \BBless signal. For any experiment in which the source is embedded in the detector, we can express the mean number of background counts $B(\delta E)$ in an energy window $\delta E$ as

\begin{equation}
B(\delta E) = b \cdot M \cdot \delta E \cdot t,
\label{eq:n-bkg}
\end{equation}
where $b$ is the background rate per unit detector mass per energy interval
(units: cts/(keV\,kg\,y)). 

Usually, $b$ is independently measured by a fit over an energy range much larger than the energy window of interest $\delta E$. The background in $\delta E$ follows a Poisson distribution with a mean value of $B(\delta E)$.
 
Eq.~\eqref{eq:n-bkg} assumes that the number of background events scales linearly with the absorber mass of the detector. We will use this simplified model for our background-fluctuation sensitivity calculations. However, other cases, most notably surface contaminations, are in fact possible wherein the background might not scale with $M$. Therefore, the final analysis of experimental data requires a detailed understanding of the physical distribution of the background sources and Monte Carlo simulations of the specific detector geometry under consideration.

We will use Eq.~\eqref{eq:nbb-full} and Eq.~\eqref{eq:n-bkg} as analytic expressions for the expected numbers of signal and background counts assuming a source-equals-detector experimental configuration, but an analogous estimation is possible for any 
detector configuration.

With the background $B(\delta E)$ defined in Eq.~\eqref{eq:n-bkg}, we can
calculate the number of counts that would represent an upward background fluctuation of a chosen significance level. For simplicity, we consider a single-bin counting experiment wherein the width of the bin is equal to the energy window $\delta E$ centered on the expected \BBless transition energy; this allows us to decouple the sensitivity calculation from the specific analysis approach used by the experiment.

In this case, the experimental sensitivity is the smallest mean signal $\widehat{S_{0}}$ that is \emph{greater than or equal to} a background fluctuation of a chosen significance level. 
If $B(\delta E)$ is large enough, the background fluctuation will be Gaussian, and the significance level can be expressed in terms of a number of Gaussian standard deviations $n_{\sigma}$. Then $\widehat{S}(\delta E)$ is given by

\begin{equation}
\widehat{S}(\delta E) = \widehat{S_{0}} \cdot f(\delta E) = n_{\sigma} \cdot \sqrt{B(\delta E)},
\label{eq:gauss-sens}
\end{equation}
where $\sigma = \sqrt{B(\delta E)}$ and $f(\delta E)$ is the fraction of signal events that fall in the energy window cut $\delta E$ around the Q-value. $f(\delta E)$ is a simple estimate of the analysis efficiency.

For a signal that is Gaussian-distributed in energy around the Q-value, the signal fraction $f(\delta E)$ is

\begin{equation}
f(\delta E) = \erf\left(\frac{\delta E}{\Delta E}\cdot\sqrt{\ln(2)}\right),
\label{eq:sig-fraction}
\end{equation}
where $\Delta E$ is the detector FWHM energy resolution. The value of $\delta E$ can be chosen to maximize the $\widehat{S}(\delta E)$-to-$\sqrt{B(\delta E)}$ ratio in the energy window of interest, which in turn optimizes the sensitivity criterion expressed by Eq.~\eqref{eq:gauss-sens}; this optimal choice corresponds to $\delta E \approx 1.2\Delta E$. It is, however, common to take $\delta E=\Delta E$.
In this case, the sensitivity differs by less than 1\% from the one calculated at the optimal cut.

By using $\widehat{S_{0}}$ and $B(\delta E)$ from Eq.~\eqref{eq:nbb-full} and \eqref{eq:n-bkg}, we obtain an expression for the background-fluctuation sensitivity of \BBless experiments in the following form:

\begin{equation}
  \widehat{\Tz\!\!}(n_{\sigma}) =\frac{\ln(2) }{n_{\sigma}} \frac{N_{A} \cdot a \cdot \eta \cdot \varepsilon}{W} \sqrt{\frac{M \cdot t}{b\cdot\delta E}}  \cdot f(\delta E).  
\label{eq:sens-exp}
\end{equation}
This equation is useful in evaluating the expected
performance of prospective experiments, as it analytically links the
experimental sensitivity with the detector parameters. Aside from the inclusion of the signal fraction, it is similar to the familiar `factor of merit' expression used within the \BBless experimental community.

For small numbers of observed events, i.e., extremely low backgrounds, the Gaussian approximation of Eq.~\eqref{eq:gauss-sens} and Eq.~\eqref{eq:sens-exp} does not provide the correct probability coverage, and therefore the meaning of the significance level is not preserved.
If $B(\delta E)$ is $\lesssim24$~counts, the Gaussian calculation of a $1\sigma$ sensitivity will differ from its Poissonian counterpart (developed below) by $10\%$ or more.

Although the Gaussian limit will possibly still be sufficient for CUORE (see Sec.~\ref{sec:CUORE}), 
a more careful calculation might be necessary in the case of a lower background or smaller exposure, or for more sensitive experiments in the future.
We therefore compute the sensitivity by assuming a Poisson distribution of the background counts.

In terms of Poisson-distributed variables, the concept expressed by Eq.~\eqref{eq:gauss-sens} becomes~\cite{Hernandez1996293}

\begin{equation}
\sum_{k=\widehat{S}(\delta E)+B(\delta E)}^{\infty}p_{B}(k) = \alpha,
\label{eq:sens-poisson}
\end{equation}
where $\alpha$ is the Poisson integrated probability that the background distribution alone will cause a given experiment to observe a total number of counts larger than $\widehat{S}(\delta E) + B(\delta E)$. Eq.~\eqref{eq:sens-poisson} can be solved only for certain values of $\alpha$ because the left-hand side is a discrete sum. To obtain a continuous representation that preserves the Poisson interpretation of Eq.~\eqref{eq:sens-poisson}, we exploit the fact that the (discrete) left-hand side of Eq.~\eqref{eq:sens-poisson} coincides with the (continuous) normalized lower incomplete gamma function $P(a,x)$ (see page 260 of Ref.~\cite{MathFunc} for details):

\begin{equation}
P(\widehat{S}(\delta E)+B(\delta E),B(\delta E)) = \alpha.
\label{eq:sens-gamma}
\end{equation}
The computation of $\widehat{S_{0}}$ from Eq.~\eqref{eq:sens-gamma}, for given values of $B(\delta E)$
and $\alpha$, is done numerically.
Once $\widehat{S_{0}}$ is computed in this way, the corresponding Poisson-regime background-fluctuation sensitivity to the half-life \Tz for neutrinoless double-beta decay is simply calculated by reversing Eq.~\eqref{eq:nbb-full}. 

For the remainder of this paper, we will use the Poisson-regime calculation based on Eq.~\eqref{eq:sens-gamma} to evaluate our \linebreak background-fluctuation sensitivity. However, to indicate the significance level with the familiar $n_{\sigma}$ notation instead of the less-intuitive $\alpha$, we will label our sensitivities with the $n_{\sigma}$ corresponding to a Gaussian upper-tail probability of $\alpha$ (for example, we will call a background-fluctuation sensitivity calculated with $\alpha=0.159$ in Eq.~\eqref{eq:sens-gamma} a `$1\sigma$ sensitivity').

\subsection{Sensitivity for Zero or Near-Zero Backgrounds}
\label{sec:zero}

It is meaningless to define sensitivity in terms of background fluctuations when $B(\delta E)\simeq0$. To develop a formula-based sensitivity calculation for the case of zero backgrounds, we consider again a single-bin counting experiment in the same way as we did for the background-fluctuation sensitivity; however, we must adopt a new method of constructing our sensitivity parameter.

To construct the zero-background sensitivity, we choose to follow the Bayesian limit-setting procedure. Instead of comparing the mean signal value $S(\delta E)$ to the mean background value $B(\delta E)$, we are now obliged to consider $S_{max}(\delta E)$, the upper limit on $S(\delta E)$ in the case that the experiment observes zero counts (i.e., no background \emph{or} signal) in $\delta E$ during its live time. $S_{\max}(\delta E)$ can be evaluated using a Bayesian calculation with a flat signal prior (see Eq.~(32.32)--(32.34) of Ref.~\cite{PDG}):

\begin{equation}
\frac{\int_{S=0}^{S_{max}(\delta E)}p_{S}(0)dS}{\int_{S=0}^{\infty}p_{S}(0)dS} = \frac{\int_{S=0}^{S_{max}(\delta E)}S^{0}e^{-S}dS}{\int_{S=0}^{\infty}S^{0}e^{-S}dS} = \frac{\text{C.L.}}{100},
\label{eq:siglimit-zerob}
\end{equation}
where $p_{S}(k)$ is the Poisson distribution $p_{\mu}(k)$ with mean $\mu=S$ and the credibility level C.L. is expressed as a percent. Eq.~\eqref{eq:siglimit-zerob} can be solved analytically for $S_{max}(\delta E)$:

\begin{equation}
S_{max}(\delta E) = S_{max} \cdot f(\delta E) = -\ln(1 - \frac{\text{C.L.}}{100}),
\label{eq:siglimit-zerob-analyt}
\end{equation}
where $S_{max}$ is the inferred upper limit on $S_{0}$. Using $S_{max}$ in place of $S_{0}$ in Eq.~\eqref{eq:nbb-full}, we obtain

\begin{equation}
\widehat{\Tz\!\!}(\text{C.L.}) = -\frac{\ln(2)}{\ln(1-\frac{\text{C.L.}}{100})} \frac{N_{A} \cdot a \cdot \eta \cdot \varepsilon}{W} M \cdot t \cdot f(\delta E).
\label{eq:sens-zerob}
\end{equation}
Depending upon the resolution of the experiment, it may be advantageous to consider a wider window than $\delta E = \Delta E$ in the zero-background case, as there is no longer the need to optimize the signal-to-background ratio; the only concern is that the window remain sufficiently narrow that the irreducible background from the $2\nu\beta\beta$ continuum remains negligible if possible.

For practical purposes, this background-free approximation becomes necessary when the background-fluctuation sensitivity in units of counts is of the order of unity or less, $\widehat{S_{0}}\lesssim1$~count. By definition, the interpretations of the zero-background sensitivity and the background-fluctuation sensitivity do not entirely coincide.

\subsection{Sensitivity with Respect to the Average Expected Limit}
\label{sec:avg-lim}

In the finite-background case, an alternative approach is to use a Monte-Carlo-based procedure to evaluate the experimental sensitivity in terms of the limit that will be set in the case that the observation is consistent with background. Following what we have done in~\cite{Andreotti:2010vj}, the method requires generating a large number of toy Monte Carlo spectra assuming zero \BBless signal in the fit window (much wider than the $\delta E = \Delta E$ window used for the background-fluctuation sensitivity, in order to utilize the available shape information in the fit). For each Monte Carlo spectrum, a binned maximum likelihood fit to the spectrum is performed and used to extract the associated Bayesian limit with a flat signal prior by integrating the posterior probability density (the same analysis technique used in~\cite{Arnaboldi:2005cg,Arnaboldi:2008ds}). Finally, the distribution of the limits calculated from the Monte Carlo spectra is constructed, and its median is taken to be the sensitivity.

This average-limit sensitivity method is, in a way, more powerful than the analytical
background-fluctuation method because it can in principle take into
account detector-dependent and experiment-specific effects, which can be difficult or sometimes
impossible to model with analytical formulas. The average-limit approach relies on analysis of statistical ensembles but lacks the simplicity offered by the analytical approach of the background-fluctuation sensitivity formulas. The two methods are, as will be shown, essentially equivalent given the same input parameters, though a minor systematic difference arises because the probability distribution of the limits is 
not symmetric and the median found with the MC does not coincide with the $\widehat{S}(\delta E)$ computed with Eq.~\eqref{eq:sens-gamma}.

For a completed experiment like Cuoricino, the experimental parameters (e.g., background rate(s) and shape(s), resolution(s), exposure) have been directly measured and are used as inputs to the Monte Carlo. The average-limit sensitivity is meaningful for a completed experiment that has not seen evidence of a signal because it provides an understanding of the experiment's real experimental prospects and whether or not it was favored by chance in the limit that it was able to set. To adapt the approach for an upcoming experiment, it is of course necessary to instead use the expected experimental parameters to generate the Monte Carlo spectra. Calculating the average-limit sensitivity in this way allows for the direct comparison of an upcoming experiment with previously reported experimental limits. The average-limit sensitivity is often considered in specific \BBless experiments; for example, the GERDA experiment reports a sensitivity calculated in essentially this manner~\cite{PhysRevD.74.092003}, although they choose to report the mean expected limit instead of the median.

\subsection{Experimental Potential to Discover \BBless}
\label{sec:discovery}

In the case of experiments like those searching for \BBless, it may be desirable to frame the experiment's capabilities in terms of discovery potential rather than sensitivity; in other words, one may wish to report the maximum \BBless half-life for which the experiment can be reasonably expected to be able to truly claim discovery of the decay. For the formulation of discovery potential, two criteria must be established: the requirement to claim discovery given a particular experimental observation, and the requirement to `reasonably expect' to obtain a particular experimental observation (in particular, one that satisfies the discovery criterion) given a particular true \BBless-signal-plus-background magnitude. The discovery potential then corresponds to the minimum true \BBless signal magnitude that would satisfy these requirements.

For sufficiently large expected backgrounds ${B(\delta E)>>1}$, the requirement to claim discovery can be straightforwardly expressed in the framework of the background-fluctuation sensitivity. When a finite background is present, it can never be entirely certain that a given observation is due to the presence of a signal, as there is always some possibility that the observation may arise from the background count distribution alone; however, the convention is that discovery may be claimed if an upward Gaussian background fluctuation of 5$\sigma$ or greater would be required to explain the observation with the background distribution alone, corresponding to a probability of $2.87\times10^{-7}$. If we state that the requirement to `reasonably expect' to be able to claim discovery is that the mean of the true (signal plus background) count distribution is at least large enough to fulfill this requirement, then the finite-background Gaussian-regime discovery potential is defined by Eq.~\eqref{eq:gauss-sens} or, equivalently, Eq.~\eqref{eq:sens-exp}, for $n_{\sigma}=5$; in essence, it is the `5$\sigma$ sensitivity.' In the Poisson regime, then, the finite-background discovery potential may be similarly considered to be the \BBless half-life that would give rise to the mean signal $\widehat{S_{0}}$ found from Eq.~\eqref{eq:sens-gamma} for $\alpha = 2.87\times10^{-7}$ and the appropriate value of $B(\delta E)$.

For very small expected background levels $B(\delta E)\simeq0$, however, we cannot continuously extrapolate the ability to claim that the experimental observation is inconsistent with the background-only hypothesis at a certain significance level; it is not possible to observe a fraction of an event, so the minimum requirement to be able to claim discovery is the observation of a single signal event. One way to ensure that the discovery potential represents the minimum possible signal for which the experiment can be reasonably expected to be able to claim discovery is to consider the true zero-background case. Unlike in the finite-background case, if the background is truly zero, the observation of a single event will satisfy the requirement to claim discovery. However, it is still necessary to set the requirement to `reasonably expect' to be able to claim discovery (i.e., observe more than zero events). This can be done by requiring that the true expected signal distribution corresponding to $\widehat{S}(\delta E)$ must yield at least a certain probability $\mathcal{P}$ of observing more than zero counts:

\begin{equation}
1 - p_{\widehat{S}(\delta E)}(0) = 1 - e^{-\widehat{S}(\delta E)} \geq \mathcal{P}.
\label{eq:disc_thres}
\end{equation}
This is mathematically equivalent to the upper limit in the case of zero observed counts that would be found from Eq.~\eqref{eq:siglimit-zerob-analyt} with a credibility level of $\mathcal{P}$.

Unlike the conventional requirement that an experimental observation must correspond to at least a 5$\sigma$ background fluctuation to claim discovery, the choice of $\mathcal{P}$ is arbitrary. It defines a flat minimum threshold in $\widehat{S}(\delta E)$ depending upon how certain one wishes to be that an experiment will observe at least one signal event in its region of interest.

For a sufficiently small expected number of background counts, depending on the choice of $\mathcal{P}$, the requirement for the expected observation to be inconsistent with background becomes less stringent than the requirement that the experiment be reasonably likely to observe any signal event at all. A simple formulation of discovery potential can therefore be established by setting two criteria:
\begin{itemize}
\item $\widehat{S}(\delta E) \geq -\ln(1 - \mathcal{P})$ and
\item $P(\widehat{S}(\delta E)+B(\delta E),B(\delta E)) \leq 2.87\times10^{-7}$,
\end{itemize}
where $P(a,x)$ is the lower normalized incomplete gamma function, as discussed in reference to Eq.~\eqref{eq:sens-gamma}.
The discovery potential curve is then defined by the minimum value of $\widehat{S}(\delta E)$ that satisfies both criteria.

Alternatively, a Monte-Carlo-based discovery potential can be constructed in an analogous manner to the average-limit sensitivity using the specific analysis mechanisms and choice of discovery criteria defined by the particular experiment. The sensitivity tools of Ref.~\cite{PhysRevD.74.092003} provide a prescription for a discovery potential calculated in this way.

\section{Limits and Sensitivities in Cuoricino}
\label{sec:Qino-to-CUORE}

Cuoricino~\cite{Arnaboldi:2004qy} achieved the greatest sensitivity of any bolometric \BBless experiment to date and served as a prototype for the CUORE experiment.  Cuoricino
took data from 2003 to 2008 in the underground facilities of the
Laboratori Nazionali del Gran Sasso (LNGS), Italy.

The Cuoricino detector consisted of 62 \TO bolometers with a total mass of 40.7~kg.
The majority of the detectors
had a size of $5\times5\times5$~cm$^3$ (790~g) and
consisted of natural $\rm TeO_2$.
The average FWHM resolution for these crystals was $6.3 \pm 2.5$~keV at 2615 keV~\cite{Andreotti:2010vj}, the nearest strong peak to the \BBless transition energy.
Their physical efficiency, which is
mostly due to the geometrical effect of beta particles escaping the detector and radiative processes, has been estimated to be $\varepsilon_{phys} = 0.874\pm0.011$~\cite{Andreotti:2010vj}.  
The full details of the crystal types present in the detector array can be found in~\cite{Andreotti:2010vj}.

Cuoricino did not see any evidence for \BBless and published a limit based on its observed spectrum, which was presented alongside an average-limit sensitivity. This sensitivity was evaluated as the median of the distribution of 90\% C.L. limits extracted from toy Monte Carlo simulations that used the measured detector parameters as inputs, and it was determined to be $\widehat{\Tz\!\!}(90\%\mbox{~ C.L.}) = 2.6 \times 10^{24}$~y.

Because of the different crystal types present in Cuoricino, if we wish to calculate a background-fluctuation sensitivity for Cuoricino to compare with this average-limit sensitivity, we need to slightly adjust the background- \linebreak fluctuation calculation presented in Sec.~\ref{sec:back-fluct} to accommodate different parameter values for the different crystal types. 
Cuoricino can be considered as the sum of virtual detectors, each representing one of the crystal types during one of the two major data-taking periods of Cuoricino's run. The detectors' total exposures, background rates after event selection, physical efficiencies, and average resolutions are reported in Ref.~\cite{Andreotti:2010vj}, subdivided by crystal type and data-taking period as appropriate. Therefore we can use these reported values to calculate both our expected signal $\widehat{S}(\delta E)$ and expected background $B(\delta E)$ as sums of the contributions from these virtual detectors, then follow the Poisson-regime background-fluctuation sensitivity procedure.
To quantitatively compare to a 90\% C.L. average-limit sensitivity, we must choose to calculate the background-fluctuation sensitivity at $1.64\sigma$ ($\alpha = 0.051$); indeed, doing so yields $\widehat{\Tz\!\!}(1.64\sigma) = 2.6 \times 10^{24}$~y, in perfect agreement with the average-limit sensitivity. 

Following previously established convention for past bolometric experiments~\cite{Ardito:2005ar,Arnaboldi200391}, we choose to report \linebreak background-fluctuation sensitivities at $1\sigma$ ($\alpha=0.159$) for upcoming experiments. For the purpose of illustration, the corresponding background-fluctuation sensitivity for Cuoricino would be $\widehat{\Tz\!\!}(1\sigma) = 4.2 \times 10^{24}$~y.

\begin{table*}[tbp]
  \centering
  \caption{Values used in the estimation of the sensitivity of CUORE-0 and CUORE. Symbols are defined in  Eq.~\eqref{eq:nbb-full}, Eq.~\eqref{eq:n-bkg}, and Eq.~\eqref{eq:gauss-sens}. See Sec.~\ref{sec:CUORE} for a discussion of the background values.}
  \small
  \begin{tabular}{c c c c c c c cc c}
    \noalign{\smallskip}
    \hline
    \noalign{\smallskip}
    \hline
    \noalign{\smallskip}
     & $a$ & $\eta$ & $\varepsilon$ & $W$ & $M$ & $\Delta E$ & $f(\Delta E)$ & $b$\\ 
    Experiment & \scriptsize{(\%)} & & \scriptsize{(\%)} & \scriptsize{(g/mol)} & \scriptsize{(kg)} & \scriptsize{(keV)} & \scriptsize{(\%)} & \scriptsize{(cts/(keV\,kg\,y))} \\ 
    \noalign{\smallskip}
    \hline
    \noalign{\smallskip}
    CUORE-0 & 34.167 & 1 & 87.4 & 159.6 & 39 & 5 & 76 & 0.05 \\ 
    CUORE &  34.167 & 1 & 87.4 & 159.6 & 741 & 5 & 76 & 0.01 \\ 
    \noalign{\smallskip}
    \hline
    \noalign{\smallskip}
    \hline
  \end{tabular}
  \label{tab:values}
\end{table*}

Although upcoming CUORE-family experiments have historically shown 1$\sigma$ background-fluctuation sensitivities, which quantitatively roughly coincide with 68\% C.L. average-limit sensitivities, some other upcoming \BBless experiments report 90\% C.L. sensitivities. To prevent confusion, it is instructive to compare 1.64$\sigma$ background-fluctuation sensitivities to 90\% C.L. average-limit sensitivities for both CUORE and CUORE-0; this comparison appears in Sec.~\ref{sec:CUORE}.

\section{Sensitivity and Discovery Potential in CUORE}
\label{sec:CUORE}

CUORE will consist of an array of 988 \TO cubic detectors, similar to
the $5\times5\times5$~cm$^3$ Cuoricino crystals described above. The total mass of the detectors will be
741~kg. The detectors will be arranged in 19 individual
towers and operated at $\sim10$~mK in the Gran Sasso underground laboratory.
The expected energy resolution FWHM of the CUORE detectors is $\Delta E
\approx 5$~keV at the \BBless transition energy, or Q-value ($\sim2528$~keV for \elem{130}{Te}~\cite{Redshaw:2009cf,Scielzo:2009nh,Rahaman2011412}). This resolution represents an improvement over that seen in Cuoricino and has
already been achieved in tests performed in the CUORE R\&D facility at
LNGS.
CUORE is expected to accumulate data for about 5 years of total
live time.  The experiment is currently being constructed and first
data-taking is scheduled for 2014.

The CUORE collaboration currently operates a single CUORE-like tower in the
former Cuoricino cryostat. This configuration,
named CUORE-0, will validate the assembly procedure and the readiness
of the background reduction measures.  The experimental parameters of
CUORE-0 and CUORE that are used in the sensitivity calculations are
summarized in Tab.~\ref{tab:values}.

\begin{figure}[tbp] 
   \centering
   \includegraphics[width=1.\columnwidth]{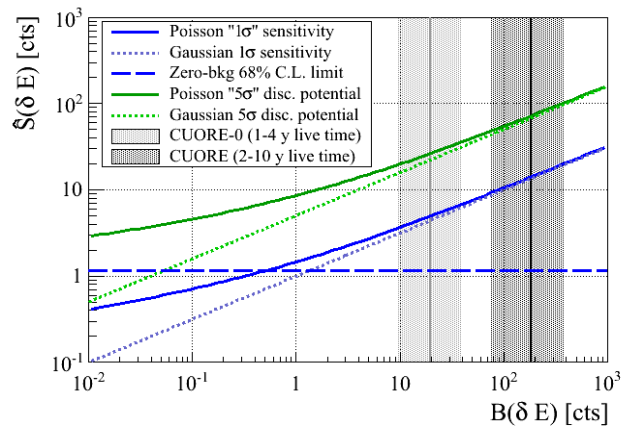} 
   \caption{Background-fluctuation sensitivity and discovery potential curves in units of counts in $\delta E$. The Poisson curve approaches the Gaussian curve at the same significance level for $B(\delta E)>>1$. For very small $B(\delta E)$, the discovery potential will follow whatever flat minimum $\widehat{S}(\delta E)$ threshold is chosen until that threshold crosses the Poisson $5\sigma$ curve. The shaded regions indicate the regimes into which CUORE-0 and CUORE are expected to fall for $\delta E = \Delta E$, given their anticipated exposures; the vertical lines indicate the values of $B(\delta E)$ corresponding to 2~y of CUORE-0 live time (with $b=0.05$~cts/(keV\,kg\,y)) and 5~y of CUORE live time (with $b=0.01$~cts/(keV\,kg\,y)), respectively.}
   \label{fig:bkg_regime}
\end{figure}

Figure~\ref{fig:bkg_regime} illustrates both the $1\sigma$ sensitivity and the $5\sigma$ discovery potential in units of signal counts in the region of interest, $\widehat{S}(\delta E)$, as a function of the number of background counts expected to be observed in the region of interest, $B(\delta E)$. For the sensitivity, curves for the zero-background (zero-count limit), small-background (Poisson background-fluctuation sensitivity), and large-background (Gaussian background-fluctuation sensitivity) regimes are all shown at equivalent significance/credibility levels. For the discovery potential, only the Poisson and Gaussian curves are shown; in this case, the zero-background criterion will depend upon the threshold criterion chosen to define the desired probability of observing at least one signal event. The background rate is the most critical parameter to assess before the calculation of the sensitivity can be carried out, as it and the exposure together determine $B(\delta E)$.

In Cuoricino, the average background counting rate in the region of interest (ROI) for \BBless decay, namely, a region centered at the Q-value and 60 keV wide, was $0.161 \pm 0.006$~cts/(keV\,kg\,y) for the $5\times5\times5$~cm$^3$ crystals\footnote{This is the background rate measured when operating the array in anticoincidence; this evaluation is extracted from the \BBless best fit~\cite{Andreotti:2010vj} and corrected for the instrumental efficiency to give the real rate.}.
An analysis of the background sources responsible for the flat background in the ROI has been performed on a partial set of statistics~\cite{Ardito:2005ar,Arnaboldi:2008ds}, following the technique and the model developed for the MiDBD experiment~\cite{Bucci2009}. The result of this analysis was the identification of three main contributions: $30 \pm 10$\% of the measured flat background in the ROI is due to multi-Compton events due to the 2615~keV gamma ray from the decay chain of \elem{232}{Th} from the contamination of the cryostat shields; $10 \pm 5$\% is due to surface contamination of the \TO crystals with \elem{238}{U} and \elem{232}{Th} (primarily degraded alphas from these chains); and $50 \pm 20$\% is ascribed to similar surface contamination of inert materials surrounding the crystals, most likely copper (other sources that could contribute are muons~\cite{Andreotti201018} and neutrons, but simulations indicate that these have only a minor effect). 

On the basis of this result, the R\&D for CUORE has pursued two major complementary avenues: one, the reduction of surface contamination, and two, the creation of an experimental setup in which potential background contributions are minimized by the selection of extremely radio-pure construction materials and the use of highly efficient shields. 
The latter activity is based mainly on standard procedures (material selection with HPGe spectroscopy, underground storage to avoid activation, evaluation of the background suppression efficiencies of the shields on the basis of Monte Carlo simulations~\cite{Bellini2010169}, etc.). However, the required surface contamination levels are extremely low, on the order of 1--10~nBq/cm$^2$. In most cases, only bolometric detectors are sufficiently sensitive to measure contaminations at this level; at this time, our understanding of these contaminations comes only from the statistics-limited data sets collected by small test detectors constructed from CUORE materials (see Ref.~\cite{Alessandria:2011vj} for the contract requirements on and measurements of the contamination levels of the crystals).

A detailed analysis of the background mitigation effort and its extrapolation to the CUORE and CUORE-0 background is out of the scope of the present paper. Here, to justify the expected background rates that will be used for the sensitivity estimations, we offer a brief summary, allowing us to perform a simple scaling to obtain the range into which we expect the CUORE-0 background rate to fall and support the conclusion that CUORE will meet its design background specification.

CUORE crystals are produced following a controlled protocol~\cite{Arnaboldi:2010fj} that is able to ensure a bulk contamination level lower than $3 \times 10^{-12}$ g/g in both \elem{238}{U} and \elem{232}{Th}. A more rigorous surface-treatment technique than that used for the Cuoricino crystals was developed; when studied with a small array of bolometric detectors, it proved to be able to reduce the surface contamination of Cuoricino crystals re-treated with this method by approximately a factor of 4~\cite{Pavan:2008zz}. The technique has now been adopted and applied in the production of the CUORE crystals, and bolometric tests have already proven its efficacy~\cite{Alessandria:2011vj}. A preliminary evaluation of the surface contaminations of the final CUORE crystals~\cite{Arnaboldi:2010fj} indicated a lower limit on the reduction with respect to the contamination seen in Cuoricino of a factor of 2; the measurement was statistics-limited, so the true reduction factor may be greater. 

In Cuoricino, a large fraction of the \BBless background was identified as due to surface contamination of the copper --- the only significant material surrounding the detectors, which are mounted in vacuum. Unfortunately, the signature of the surface contamination of the copper is extremely weak when compared to other contributions, as the background ascribed to the copper contamination is a flat continuum that can be easily observed only in the peakless 3--4 MeV region of the spectrum~\cite{Bucci2009,Pavan:2008zz}. Extensive efforts have been dedicated to the study of different treatment procedures able to reduce the copper surface contamination~\cite{Alessandria2013}; in the end, a technique that proved to be capable of reducing the copper surface contamination by at least a factor of 2 as compared with that observed in Cuoricino has been selected by the collaboration as the baseline for the CUORE copper treatment. 

Based on the above-reported considerations, we define a conservative case wherein we assume that the specific contaminations of the CUORE copper and crystals have both been reduced by a factor of 2 relative to Cuoricino.
CUORE-0 will be able to measure the level of radiopurity achieved with the chosen surface treatment.

CUORE-0 will consist of CUORE crystals mounted in CUORE-style frames as a single tower. Because of this geometry, which is similar to that of Cuoricino, the contamination reduction factors reported above scale almost directly to the background we expect to observe in the ROI. The total amount of copper facing the crystals will be only slightly reduced with respect to Cuoricino, but its surface will be treated with the new procedure studied for CUORE. CUORE-0 will be assembled in the Cuoricino cryostat, so the gamma background from contamination in the cryostat shields will remain approximately the same as in Cuoricino.
We consider that the irreducible background for CUORE-0 comes from the 2615~keV \elem{208}{Tl} line due to \elem{232}{Th} contaminations in the cryostat, in the case that all other background sources (i.e., surface contaminations) have been rendered negligible; this would imply a lower limit of $\sim0.05$~cts/(keV\,kg\,y) on the expected background in CUORE-0. Similarly, an upper limit of 0.11~cts/(keV\,kg\,y) follows from scaling the Cuoricino background in the conservative case, described above, of a factor of 2 improvement in crystal and copper contamination.

\begin{figure}[tbp] 
   \centering
   \includegraphics[width=1.\columnwidth]{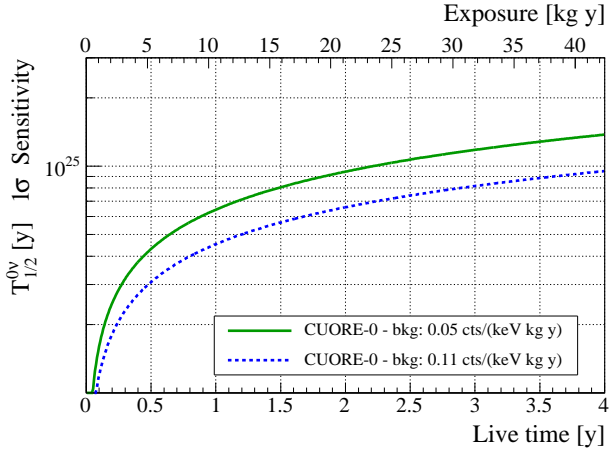} 
   \caption{CUORE-0 background-fluctuation sensitivity at $1\sigma$ for two different values
     of the background rate in the region of interest, $0.05$~cts/(keV\,kg\,y) (solid line) and $0.11$~cts/(keV\,kg\,y) (dotted line), representing the range
     into which the CUORE-0 background is expected to fall. }
   \label{fig:CUORE0-sens-Tz}
\end{figure}

A plot of the expected $1\sigma$ background-fluctuation sensitivity of CUORE-0 as a function of live time in these two bounding cases is shown in Fig.~\ref{fig:CUORE0-sens-Tz}. Tab.~\ref{tab:sens-CUORE0} provides a quantitative comparison among $1\sigma$ background-fluctuation sensitivities (as shown in Fig.~\ref{fig:CUORE0-sens-Tz}), $1.64\sigma$ background-fluctuation sensitivities, 90\%~C.L. average-limit sensitivities, and $5\sigma$ discovery potentials for CUORE-0 at several representative live times. The anticipated total live time of CUORE-0 is approximately two years; for this live time at the $0.05$~cts/(keV\,kg\,y) background level, $B(\delta E)\sim20$~cts, meaning that the Poisson-regime calculation is necessary. The pale shaded region in Fig.~\ref{fig:bkg_regime} illustrates where the CUORE-0 live time range considered in Tab.~\ref{tab:sens-CUORE0} lies with respect to the statistical regime of the sensitivity calculations for the $0.05$~cts/(keV\,kg\,y) background level.

\begin{table*}[tdp]
\centering
\caption{Several estimators of the experimental capabilities of CUORE-0 under different background estimations after one, two, and four years of live time. The boldfaced column corresponds to the anticipated total live time of two years. The background-fluctuation half-life sensitivities at $1\sigma$ are the official sensitivity values reported by the collaboration. $1.64\sigma$ background-fluctuation sensitivities and 90\%~C.L. average-limit sensitivities, in italics, are provided to illustrate the similarity of these two values to one another. The $5\sigma$ discovery potentials for $\mathcal{P} = 0.90$ are also given.}
\small
\begin{tabular}{c@{\hspace{10mm}}c@{\hspace{10mm}}c@{\hspace{10mm}}c@{\hspace{5mm}}c@{\hspace{5mm}}c@{\hspace{5mm}}c}
\noalign{\smallskip}
\hline
\noalign{\smallskip}
\hline
\noalign{\smallskip}
 & & & \multicolumn{3}{c}{half-life sensitivity} \\
$b$ & $\Delta E$ & Method & \multicolumn{3}{c}{\scriptsize{($10^{25}$ y)}} \\ 
\scriptsize{(cts/(keV\,kg\,y))} & \scriptsize{(keV)} & \scriptsize{(sig./cred. level)} & \footnotesize{1 y} & \footnotesize{\textbf{2 y}} & \footnotesize{4 y} \\
\noalign{\smallskip}
\hline
\noalign{\smallskip}
0.11 & 5 & $1\sigma$ & 0.45 & \textbf{0.66} & 0.95\\
 & & $\mathit{1.64\sigma}$ & \textit{0.28} & \textbf{\textit{0.40}} & \textit{0.58}\\
 & & \textit{90\%~C.L.} & \textit{0.29} & \textbf{\textit{0.41}} & \textit{0.59}\\
 & & $5\sigma$ & 0.085 & \textbf{0.13} & 0.18\\
 0.05 & 5 & $1\sigma$ & 0.64 & \textbf{0.94} & 1.4\\
 & & $\mathit{1.64\sigma}$ & \textit{0.39} & \textbf{\textit{0.58}} & \textit{0.84}\\
 & & \textit{90\%~C.L.} & \textit{0.39} & \textbf{\textit{0.59}} & \textit{0.83}\\ 
 & & $5\sigma$ & 0.12 & \textbf{0.18} & 0.26\\
\noalign{\smallskip}
\hline
\noalign{\smallskip}
\hline
\end{tabular} 
\label{tab:sens-CUORE0}
\end{table*}

CUORE, in addition to the new crystals and frames already present in CUORE-0, will be assembled as a 19-tower array in a newly constructed cryostat. The change in detector geometry will have two effects. First, the large, close-packed array will enable significant improvement in the anticoincidence analysis, further reducing crystal-related backgrounds.
 Second, the fraction of the total crystal surface area facing the outer copper shields will be reduced by approximately a factor of 3. In addition to these considerations, the new cryostat will contain thicker lead shielding and be constructed of cleaner material, which is expected to result in a gamma background approximately an order of magnitude lower than that in the Cuoricino cryostat. Based on the above considerations and the Cuoricino results, CUORE is expected to achieve its design background value of~$0.01$~cts/(keV\,kg\,y).

\begin{figure}[tbp] 
   \centering
   \includegraphics[width=1.\columnwidth]{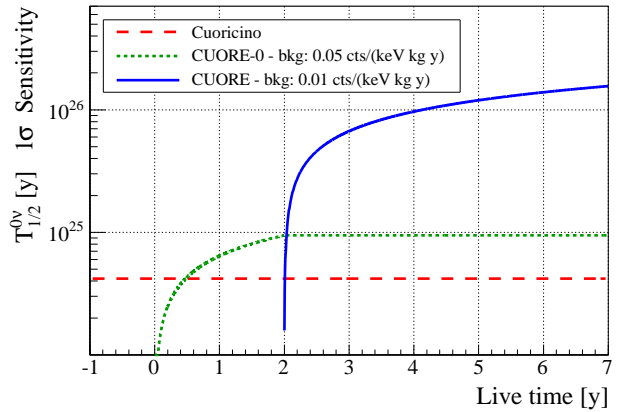} 
   \caption{Background-fluctuation sensitivities at $1\sigma$ for the CUORE-0 (dotted line) and CUORE (solid line)
     experiments, calculated from Eq.~\eqref{eq:sens-gamma} and Eq.~\eqref{eq:nbb-full} with the experimental parameters shown in Tab.~\ref{tab:values}. The Cuoricino 1$\sigma$ background-fluctuation sensitivity calculation (dashed line) is discussed in Sec.~\ref{sec:Qino-to-CUORE}.}
   \label{fig:combined-sens-Tz}
\end{figure}

An overview of the $1\sigma$ background-fluctuation sensitivities of the Cuoricino, CUORE-0, and CUORE \TO bolometric experiments is
shown in Fig.~\ref{fig:combined-sens-Tz}. The Cuoricino $1\sigma$ sensitivity calculated in Sec.~\ref{sec:Qino-to-CUORE} is shown for reference. A 1$\sigma$ half-life sensitivity close to $10^{25}$~years is expected from 2 years' live time of CUORE-0. Once CUORE starts data-taking, another order of magnitude improvement
in sensitivity is expected in another two years.

\begin{figure}[tbp] 
   \centering
   \includegraphics[width=1.\columnwidth]{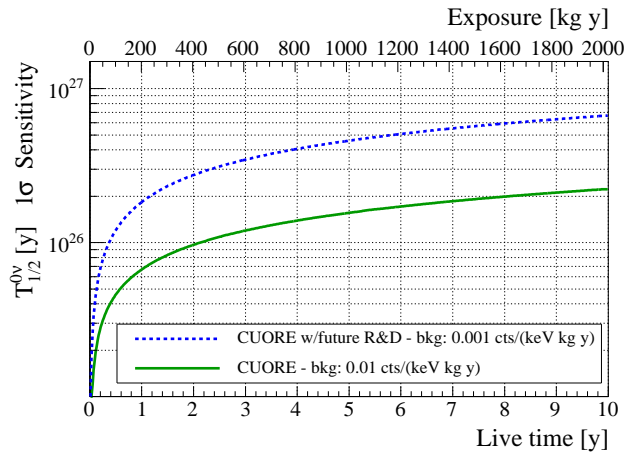} 
   \caption{Background-fluctuation sensitivity of the CUORE experiment at $1\sigma$ (solid line) for the design goal background level. The sensitivity for an order-of-magnitude improvement over the baseline background is also shown (dotted line). 
     }
   \label{fig:Q-sens-bkg-Tz}
\end{figure}

A plot of the CUORE experiment's sensitivity as a function of the live time and exposure 
is shown in Fig.~\ref{fig:Q-sens-bkg-Tz}. Tab.~\ref{tab:sens-CUORE} provides a quantitative comparison among $1\sigma$ background-fluctuation sensitivities (as shown in Fig.~\ref{fig:Q-sens-bkg-Tz}), $1.64\sigma$ background-fluctuation sensitivities, 90\%~C.L. average-limit sensitivities, and $5\sigma$ discovery potentials for CUORE at several representative live times. The anticipated total live time of CUORE is approximately five years; for this live time at the design goal background level, $B(\delta E)\sim190$~cts, meaning that the Gaussian approximation would still be valid in this case. The sensitivity values we show in this paper nevertheless differ from those previously reported by the experiment~\cite{Arnaboldi:2002du,Ardito:2005ar} by about $25\%$. This difference can be attributed to the inclusion of the signal fraction $f(\delta E)$, which was not previously considered. The dark shaded region in Fig.~\ref{fig:bkg_regime} illustrates where the CUORE live time range considered in Tab.~\ref{tab:sens-CUORE} lies with respect to the statistical regime of the sensitivity calculations for the $0.01$~cts/(keV\,kg\,y) background level.

\begin{table*}[tdp]
\centering
\caption{Several estimators of the experimental capabilities of CUORE after two, five, and ten years of live time. The boldfaced column corresponds to the anticipated total live time of five years. The values are reported for the design goal background level, as well as for an order-of-magnitude improvement over the design goal. The background-fluctuation half-life sensitivities at $1\sigma$ are the official sensitivity values reported by the collaboration. $1.64\sigma$ background-fluctuation sensitivities and 90\%~C.L. average-limit sensitivities, in italics, are provided to illustrate the similarity of these two values to one another. The 5$\sigma$ discovery potentials for $\mathcal{P} = 0.90$ are also given.}
\small
\begin{tabular}{c@{\hspace{10mm}}c@{\hspace{10mm}}c@{\hspace{10mm}}c@{\hspace{5mm}}c@{\hspace{5mm}}c@{\hspace{5mm}}c}
\noalign{\smallskip}
\hline
\noalign{\smallskip}
\hline
\noalign{\smallskip}
 & & & \multicolumn{3}{c}{half-life sensitivity} \\
$b$ & $\Delta E$ & Method & \multicolumn{3}{c}{\scriptsize{($10^{26}$ y)}} \\ 
\scriptsize{(cts/(keV\,kg\,y))} & \scriptsize{(keV)} & \scriptsize{(sig./cred. level)} & \footnotesize{2 y} & \footnotesize{\textbf{5 y}} & \footnotesize{10 y} \\
\noalign{\smallskip}
\hline
\noalign{\smallskip}
0.01 & 5 & $1\sigma$ & 0.97 & \textbf{1.6} & 2.2\\
 & & $\mathit{1.64\sigma}$ & \textit{0.59} & \textbf{\textit{0.95}} & \textit{1.4}\\
 & & \textit{90\%~C.L.} & \textit{0.59} & \textbf{\textit{0.97}} & \textit{1.4}\\
 & & $5\sigma$ & 0.19 & \textbf{0.30} & 0.44\\
 0.001 & 5 & $1\sigma$ & 2.7 & \textbf{4.6} & 6.7\\
 & & $\mathit{1.64\sigma}$ & \textit{1.7} & \textbf{\textit{2.8}} & \textit{4.1}\\
 & & \textit{90\%~C.L.} & \textit{1.6} & \textbf{\textit{2.8}} & \textit{4.2}\\
 & & $5\sigma$ & 0.50 & \textbf{0.86} & 1.3\\
\noalign{\smallskip}
\hline
\noalign{\smallskip}
\hline
\end{tabular} 
\label{tab:sens-CUORE}
\end{table*}

While it is unlikely that CUORE itself will reach a background rate of 0.001~cts/(keV\,kg\,y) or below, R\&D activities are already underway pursuing ideas for further reduction of the background in a possible future experiment. Techniques for active background rejection are being investigated~\cite{Foggetta:2009ds,Foggetta:2011nk}) that could provide substantial reduction of the background. Sensitivities for a scenario with 0.001 cts/(keV\,kg\,y) in a CUORE-like experiment are given in Fig.~\ref{fig:Q-sens-bkg-Tz} and Tab.~\ref{tab:sens-CUORE}.

\section{Comparison with the Claim in \elem{76}{Ge}}

It is interesting to compare the CUORE-0 and CUORE sensitivities with
the claim for observation of \BBless in
\elem{76}{Ge}~\cite{KlapdorKleingrothaus:2001ke,KlapdorKleingrothaus:2004ge,KlapdorKleingrothaus:2006ff}. The authors of this claim have reported several different values for the half-life of \elem{76}{Ge}, depending upon the specifics of the analysis; the longest of these, and thus the one requiring the greatest sensitivity to probe, is $\Tz(\elem{76}{\text{Ge}}) = 2.23^{+0.44}_{-0.31}\times 10^{25}$~y~\cite{KlapdorKleingrothaus:2006ff}.  From Eq.~\eqref{eq:mnu-NME}, it follows that

$$
\Tz(\elem{130}{\text{Te}}) =
\frac{F_{N}(\elem{76}{\text{Ge}})}{F_{N}(\elem{130}{\text{Te}})}\cdot\Tz(\elem{76}{\text{Ge}}).
$$
However, correlations between the $F_{N}$ calculations for the two nuclides should be taken into account. A method of treating NME uncertainties based on a previous iteration of the QRPA-T calculations is suggested, and shown to be roughly consistent with the QRPA-J and ISM calculations, in~\cite{Faessler:2008xj}. Although the values have not been updated to utilize the most recent QRPA-T calculations, the authors argue in a recent addendum to the original article that they remain a valid estimate of the spread of NME calculations~\cite{Faessler:2013hz}.
Following this method and applying the phase space factors reported in~\cite{Kotila:2012zza} (with a correction for different input parameters~\cite{Cowell:2005jw}), the expected $1\sigma$ range of $\Tz(\elem{130}{\text{Te}})$ is $\mbox{(0.49 -- 1.0)} \times 10^{25}$~y (including the $1\sigma$ uncertainty on the $^{76}$Ge claim as done in~\cite{Faessler:2008xj}).

\begin{figure}[tbp] 
   \centering
   \includegraphics[width=1.\columnwidth]{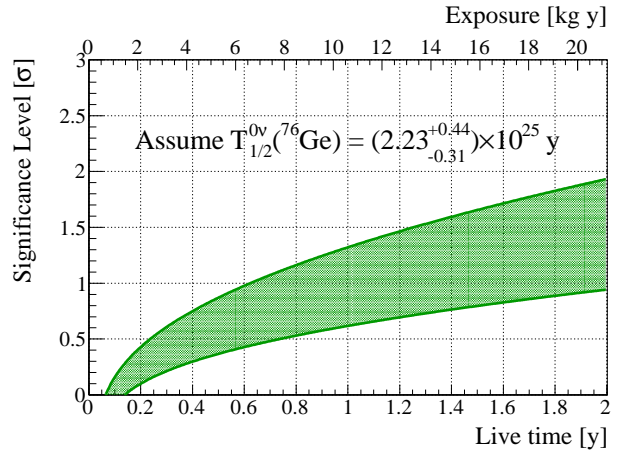} 
   \caption{Significance level at which CUORE-0 can observe a signal corresponding to the \elem{76}{Ge} claim, assuming the best expected background of 0.05~cts/(kev\,kg\,y). The width of the band accounts for both the $1\sigma$ uncertainty on the $^{76}$Ge claim and the $1\sigma$ range of QRPA-T NMEs calculated in~\cite{Faessler:2008xj}, but it is by far dominated by the NME spread.}
   \label{fig:CUORE0-vs-Klapdor}
\end{figure}

The mathematical framework of the background- \linebreak fluctuation sensitivity calculation can be inverted to determine the magnitude of the mean signal in terms of $n_{\sigma}$ that an assumed `true' half-life value will produce in an experiment. Fig.~\ref{fig:CUORE0-vs-Klapdor} shows the $n_{\sigma}$ significance level at which \mbox{CUORE-0} can probe the \elem{76}{Ge} claim as it accrues statistics over its anticipated live time. The band is bounded by curves corresponding to the maximum and minimum $\Tz(\elem{130}{\text{Te}})$ of the range given above. As can be deduced from the plot, CUORE-0 will achieve at least a 1$\sigma$ sensitivity to any signal within the expected $1\sigma$ range of $\Tz(\elem{130}{\text{Te}})$ within two years.

Thanks to the increased size and lower background, if the \elem{130}{Te} \BBless half-life indeed falls in the $1\sigma$ range implied by the claim in \elem{76}{Ge}, CUORE
will already be able to achieve a 5$\sigma$ expected signal above background within about six months.

\section{Conclusions}

In recent years, experimenters have made great strides in the search for neutrinoless double-beta decay, a discovery which would establish the Majorana nature of the neutrino and have far-reaching ramifications in physics. Next-generation \BBless experiments like CUORE have two primary goals: to test the claim of observation of \BBless in \elem{76}{Ge}, and to begin to probe effective neutrino masses of m$_{\beta\beta}~\leq$~50~meV (commonly referred to as the `inverted hierarchy region' of the neutrino mass phase space). We have investigated the expected performance of CUORE, allowing evaluation of its ability to meet these two goals.

We developed two different approaches to calculating experimental sensitivity: the background-fluctuation sensitivity and the average-limit sensitivity. The background-fluctuation sensitivity characterizes the performance of the experiment in terms of the expected background fluctuations, while the average-limit sensitivity is the average limit that the experiment expects to set in the case that there is no signal to find. The average-limit sensitivity directly compares to previously reported limits by construction, while the background-fluctuation sensitivity can be straightforwardly extended to express an experiment's capabilities in terms of discovery potential. The two methods produce quantitatively similar results, and one can consider the background-fluctuation sensitivity as an approximation of the average-limit sensitivity if the significance/credibility levels of the two methods are properly chosen to coincide.

Tab.~\ref{tab:summary} contains a summary of 1$\sigma$ background-fluctuation sensitivities to the neutrino Majorana mass according to
different NME calculations, assuming that the exchange of a light Majorana neutrino is the dominant \BBless mechanism, as discussed in Sec.~\ref{sec:dbdphys}. These values are considered the official sensitivity values for CUORE-family experiments.
During its run, CUORE will fully explore the \elem{130}{Te} \BBless half-life range corresponding to the the claim of observation of \BBless in \elem{76}{Ge}.

For illustrative purposes, Tab.~\ref{tab:summary} also shows the limiting ``zero-background'' case for both CUORE-0 and CUORE, calculated with a window of $\delta E = 2.5\Delta E$. The calculation is performed at 68\%~C.L. so that the values can be considered as zero-background extrapolations of the finite-background $1\sigma$ background-fluctuation sensitivities. To achieve this sensitivity, CUORE-0 would require $b\lesssim5\times10^{-4}$~cts/(keV\,kg\,y); CUORE would require $b\lesssim1\times10^{-5}$~cts/(keV\,kg\,y), three orders of magnitude better than the baseline background rate.
 
\begin{table*}[tbp]
\centering
\caption{Summary table of expected parameters and $1\sigma$ background-fluctuation sensitivity in half-life and effective Majorana neutrino mass. The different values of $m_{\beta\beta}$ depend on the different NME calculations; see Sec.~\ref{sec:dbdphys} and Tab.~\ref{tab:nmev}. Zero-background sensitivities for a window of $\delta E = 2.5\Delta E$, in italics, are also provided as an estimation of the ideal limit of the detectors' capabilities; they are presented at 68\%~C.L. so that they can be considered as approximate extrapolations of the $1\sigma$ background-fluctuation sensitivities.}
\small
\begin{tabular}{l@{\hspace{8mm}}c@{\hspace{8mm}}c@{\hspace{8mm}}c@{\hspace{8mm}}c@{\hspace{4mm}}c@{\hspace{4mm}}c@{\hspace{4mm}}c@{\hspace{4mm}}c@{\hspace{4mm}}c}
  \noalign{\smallskip}
  \hline
  \noalign{\smallskip}
  \hline
  \noalign{\smallskip}
  & & & & \multicolumn{6}{c}{\nmm}\\
  & $t$ & $b$ & $\widehat{\Tz\!\!}(1\sigma)$ & \multicolumn{6}{c}{\scriptsize{(meV)}}\\
  \multicolumn{1}{c}{Setup} & \scriptsize{(y)} & \scriptsize{(cts/(keV\,kg\,y))} & \scriptsize{(y)} & \footnotesize{QRPA-T} & \footnotesize{QRPA-J} & \footnotesize{ISM} & \footnotesize{IBM-2} & \footnotesize{PHFB} & \footnotesize{GCM} \\
  \noalign{\smallskip}
  \hline
  \noalign{\smallskip}
  CUORE-0 & 2 & 0.05 & 9.4$\times10^{24}$ & 160--280 & 170--290 & 340--420 & 190--210 & 170--300 & 170 \\
  \multicolumn{3}{r}{\textit{zero-bkg. case at 68\% C.L.:}} & $\mathit{5.3\times10^{25}}$ & \textit{68--120} & \textit{73--120} & \textit{140--180} & \textit{79--91} & \textit{74--130} & \textit{74} \\
  CUORE baseline & 5 & 0.01& 1.6$\times10^{26}$ & 40--69 & 42--71 & 83--100 & 46--53 & 43--74 & 43 \\
  \multicolumn{3}{r}{\textit{zero-bkg. case at 68\% C.L.:}} & $\mathit{2.5\times10^{27}}$ & \textit{9.9--17} & \textit{11--18} & \textit{21--26} & \textit{12-13} & \textit{11-18} & \textit{11} \\
  \noalign{\smallskip}
  \hline
  \noalign{\smallskip}
  \hline
\end{tabular} 
\label{tab:summary}
\end{table*}

\begin{figure*}[tbp]
  \centering
  \subfigure[]{ \includegraphics[width=1.\columnwidth]{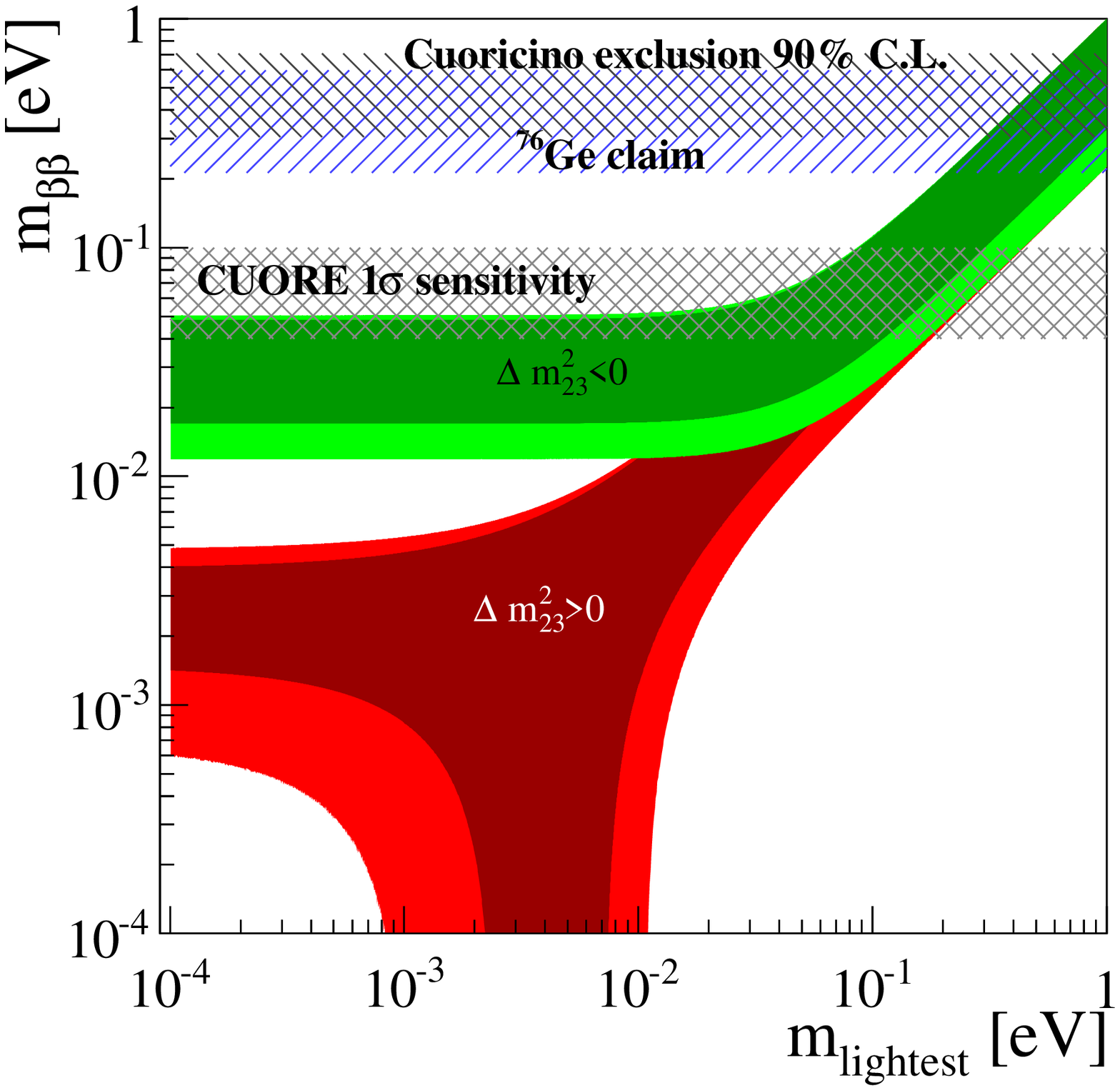} }
  \subfigure[]{ \includegraphics[width=1.\columnwidth]{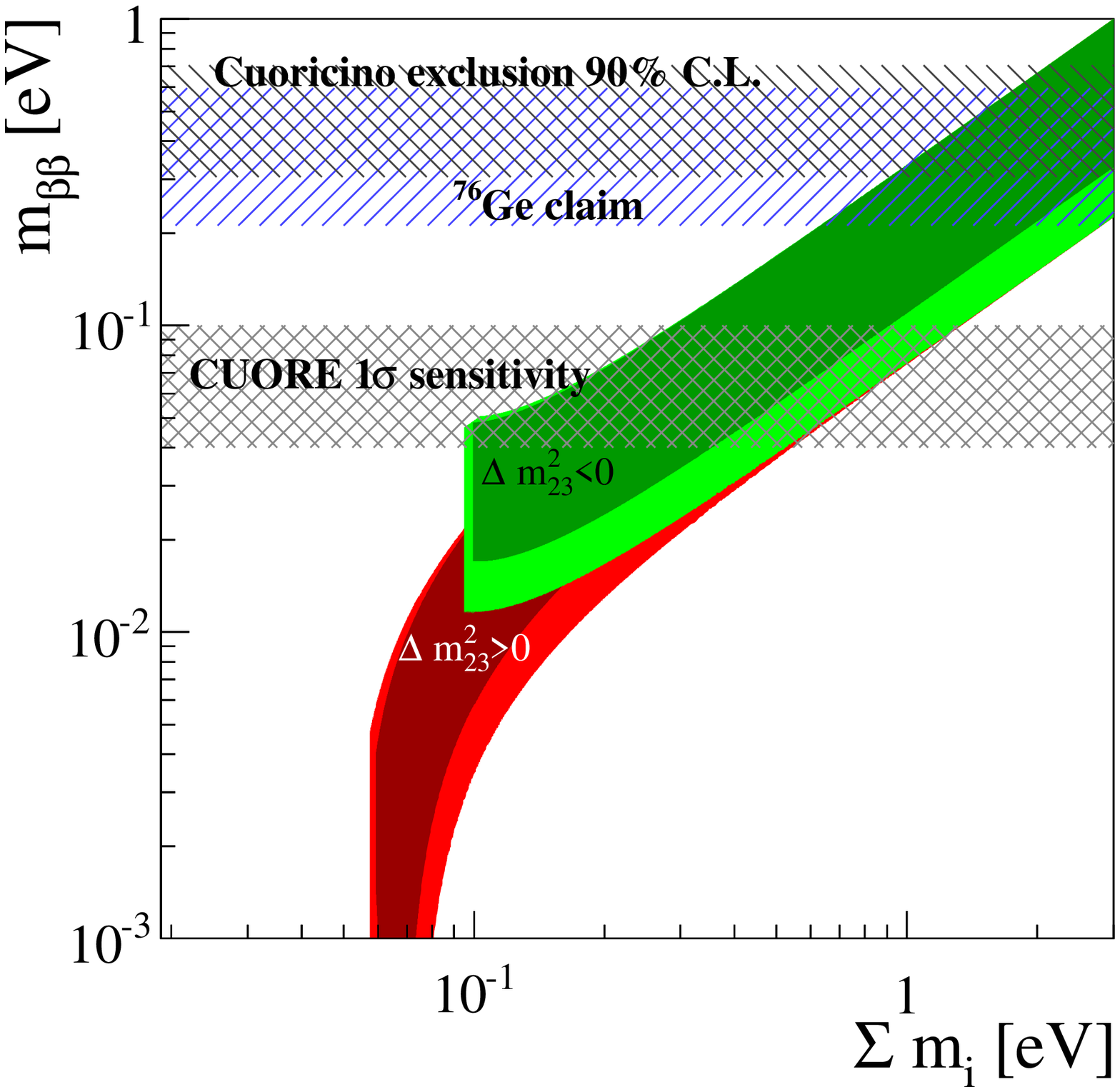} }

  \caption{The Cuoricino result and the expected CUORE $1\sigma$ background-fluctuation sensitivity overlaid on plots
    that show the bands preferred by neutrino oscillation data
    (inner bands represent best-fit data; outer bands represent data allowing
    3$\sigma$ errors)~\cite{PhysRevD.86.073012}. Both normal ($\Delta m^2_{23}>0$) and
    inverted ($\Delta m^2_{23}<0$) neutrino mass hierarchies are shown.
    (a) The coordinate plane represents the parameter space of \nmm~and
    $m_{lightest}$, following the plotting convention
    of~\cite{strumia_implications_2005}.
    (b) The coordinate plane represents the parameter space of \nmm~and
    $\Sigma m_{i}$, following the plotting convention
    of~\cite{Fogli:2008ig}.  The width of the CUORE
    band is determined by the maximum and minimum values of \nmm~obtained
    from the six NME calculations considered in this work.}
   \label{fig:sum-vs-mbb}
\end{figure*}

In Fig.~\ref{fig:sum-vs-mbb}, the expected sensitivity of
CUORE is compared with the preferred values of the neutrino mass
parameters obtained from neutrino oscillation experiments. The
sensitivity of CUORE will allow the investigation of the upper region
of the effective Majorana neutrino mass phase space corresponding to the inverted hierarchy of neutrino masses.

\section*{Acknowledgments}
The CUORE Collaboration thanks the Directors and Staff of the Laboratori Nazionali del Gran Sasso and the technical staffs of our Laboratories. This work was supported by the Istituto Nazionale di Fisica Nucleare (INFN); the Director, Office of Science, of the U.S. Department of Energy under Contract Nos. DE-AC02-05CH11231 and DE-AC52-07NA27344; the DOE Office of Nuclear Physics under Contract Nos. DE-FG02-08ER41551 and DEFG03-00ER41138; the National Science Foundation under Grant Nos. NSF-PHY-0605119, NSF-PHY-0500337, NSF-PHY-0855314, and NSF-PHY-0902171; the Alfred P. Sloan Foundation; and the University of Wisconsin Foundation.

\bibliographystyle{model1-num-names}

\bibliography{biblio}

\end{document}